\documentclass[]{article}

\usepackage[a4paper, left=0.6in, right=0.6in, top=0.6in, bottom=0.6in, footskip=.25in]{geometry}
\usepackage{authblk}
\usepackage{graphicx}

\usepackage{multirow}%
\usepackage{amsmath,amssymb,amsfonts}%
\usepackage{amsthm}%
\usepackage{mathrsfs}%

\title{Shape Morphing Metamaterials}

\author[1,2]{Krzysztof K. Dudek}
\author[2]{Muamer Kadic}
\author[3]{Corentin Coulais}
\author[4]{Katia Bertoldi}

\affil[1]{Institute of Physics, University of Zielona Gora, ul. Szafrana 4a, Zielona Gora 65-069, Poland}

\affil[2]{SUPMICROTECH, Universit\'{e} de Franche-Comt\'{e}, CNRS, Institut FEMTO-ST, F-25000 Besançon, France}

\affil[3]{Institute of Physics, Universiteit van Amsterdam, 1098 XH Amsterdam, The Netherlands}

\affil[4]{School of Engineering and Applied Sciences, Harvard University, Cambridge, MA 02138 USA}

\date{}

\begin{document}

\maketitle

\begin{abstract}
	Mechanical metamaterials leverage geometric design to achieve unconventional properties, such as high strength at low density, efficient wave guiding, and complex shape morphing. The ability to control shape changes builds on the complex relationship between geometry and nonlinear mechanics, and opens new possibilities for disruptive technologies across diverse fields, including wearable devices, medical technology, robotics, and beyond.
	In this review of shape-morphing metamaterials, we examine the current state of the field and propose a unified classification system for the mechanisms involved, as well as the design principles underlying them. Specifically, we explore two main categories of unit cells—those that exploit structural anisotropy or internal rotations—and two potential approaches to tessellating these cells: based on kinematic compatibility or geometric frustration. We conclude by discussing the available design tools and highlighting emerging challenges in the development of shape-morphing metamaterials.
\end{abstract}

\section{Introduction}
All around us, things change shape, often in a highly controlled fashion; think of the blooming of a flower, the intricate folding of a protein, the expansion of a deployable shelter, or the purposeful movement of a walking robot. The ability to program specific shape changes into structures opens up numerous compelling applications in deployable and reconfigurable materials and structures \cite{Filipov_Paulino_2015_origami_folding, Davidson_Lewis_2020_light_resp_liq_cryst, Choi_Dudte_Maha_2021_Phys_Rev_Res, Silverberg_Cohen_Science_2014}, robotics \cite{Faber_Arrieta_Studart_Science_2018, Yang_Whitesides_Adv_Mater_2015, Kim_Byun_2019}, and biomedical devices \cite{Zadpoor_Malda_2017, MacQueen_Mahadevan_heart_2018}. Mechanical metamaterials \cite{Bertoldi_van_Hecke_review_2017, Craster_2023_review}---carefully structured materials with mechanical properties governed by structure, rather than
composition---present a promising platform for realizing such controlled shape changes. While initial efforts in the field focused on achieving unconventional elastic properties such as auxetic response \cite{Lakes1987}, more recently, the research community has been exploring opportunities to embed nonlinear, non-periodic deformations into mechanical metamaterials, leading to targeted shape changes.

In this review, we concentrate on the specific topic of shape-changing metamaterials and refer readers to numerous other reviews on mechanical metamaterials for a broader perspective~\cite{Bertoldi_van_Hecke_review_2017, Zadpoor_review_2016, Greer_review_2022, Mingchao_review_Soft_Science_2023, Craster_2023_review}. Our goal is to unify recent literature, to offer a pedagogical introduction for researchers interested in shape-changing metamaterials, to highlight the field's main achievements, and to identify the remaining challenges. We address the following questions: What types of shape changes can metamaterials achieve? What are the different classes of shape-changing metamaterials? What principles enable the design of target shape changes? 

Since shape-changing metamaterials require both careful unit cell design and robust strategies for their patterning, we first present the working principles for unit cells used in shape-changing metamaterials and then survey the various strategies for assembling these unit cells into metamaterials capable of achieving target shape changes. Finally, we briefly discuss the design tools used to solve the inverse problem of identifying suitable metamaterial geometries for specific target shape changes. We conclude by providing an overview of the challenges and opportunities in the field of shape-changing metamaterials.

\section{Unit cell design principles}\label{unitcells}

The unit cells in shape-changing metamaterials proposed to date can be classified into two main categories: those based on anisotropic responses and those based on internal rotations. Anisotropic unit cells are typically made with elongated structural elements, like fibers and tubes, where axial deformations are energetically unfavorable. In contrast, rotation-based unit cells consist of stiff structural elements connected by slender joints, and their deformation is primarily driven by rotations about these joints, leading to significant changes in volume.

\subsection{Unit cells based on anisotropic mechanism}

\begin{figure}
	\centering
	\includegraphics[width=0.98\textwidth]{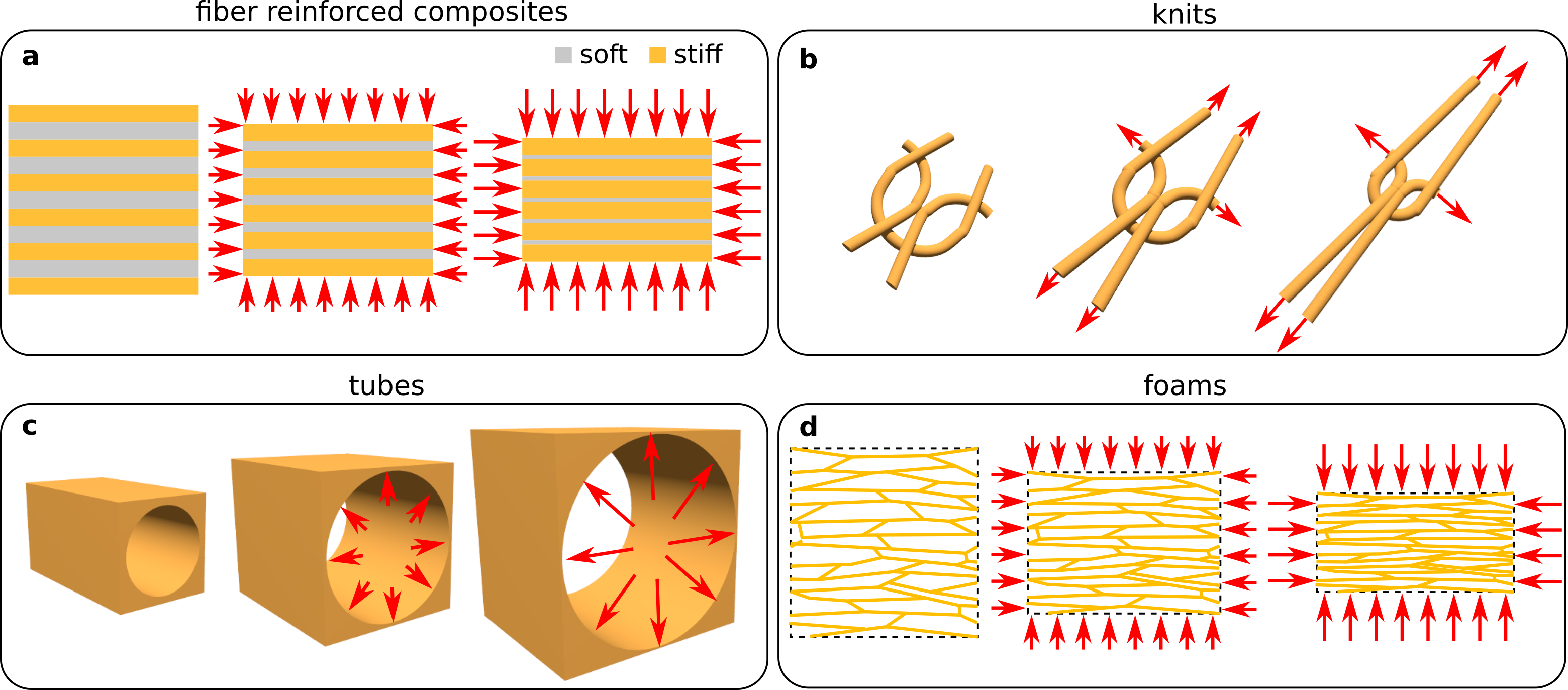}
	\caption{{\bf Unit cells based on anisotropic mechanism}. 
		%Classification of different types of anisotropy-based unit cells. 
		Panels show anisotropic unit cells made from {\bf a} fiber-reinforced composites, {\bf b} knits (adapted from \cite{Maziz_Sci_Adv_2017}), {\bf c} pneumatic tubes and {\bf d} foams that can be used to assemble larger anisotropy-based metamaterials. Red arrows on all panels indicate the applied loading.
	}
	\label{fig1}
\end{figure}

Nature	provides many examples of thin objects for which the anisotropy induced by the	orientations	of	internal	fibers	results in complex shape changes. These include the hygroscopic	actuation	of	wheat	awns \cite{Fratzl_Elbaum_Faraday_Discus_2008}, the	formation of the chiral	shape	of	some	seed	pods \cite{Armon_Sharon_Science_2011}, and the opening of pine cones~\cite{Dawson1997}. Through bioinspiration and biomimicry, many designs of anisotropic unit cells have been proposed. As illustrated in Fig. \ref{fig1}, four primary categories of anisotropic unit cells have been utilized to create shape-morphing metamaterials: fiber-elastomer composites, knits, tubes, and anisotropic foams. We briefly discuss these categories below.

\noindent {\it Fiber-elastomer composites. ---}
Fiber-elastomer composites consist of stiff fibers embedded in a soft matrix \cite{MICHALEK_Cohen_2019,Studart_NatComm2013,Schaffner_NatComm2018,Gladman_Lewis_2016}. 
Although these composites typically undergo large deformations, their key behavioral characteristics can be captured by a simple orthotropic linear elastic model. In two dimensions (under plane stress conditions), this model relates  the strain, $\varepsilon_{ij}$, and  stress, $\sigma_{ij}$,  components as 

\begin{equation}
\begin{bmatrix}
\varepsilon_{11}\\
\varepsilon_{22}\\
\varepsilon_{12}
\end{bmatrix}
= \begin{bmatrix}
\frac{1}{E_{1}} & -\frac{\nu_{12}}{E_{1}} & 0\\
-\frac{\nu_{12}}{E_{1}} & \frac{1}{E_{2}} & 0 \\
0 & 0 & \frac{2}{G}
\end{bmatrix}
\begin{bmatrix}
\sigma_{11}\\
\sigma_{22}\\
\sigma_{12}
\end{bmatrix},
\label{Hookes_law_1}
\end{equation}
where  $E_1$ and $E_2$ denote the Young's modulus along the $x_1$ and $x_2$ directions, $\nu_{12}$ represents the Poisson's ratio and $G$ is the shear modulus.
For a composite with fibers oriented along the  $x_1$ direction (see Fig. \ref{fig1}a) $E_1\gg E_2$, since the fibers resist deformations along their axis. It follows that a simple hydrostatic loading, $\sigma_{ij}=P\delta_{ij}$, leads to deformations perpendicular to the fibers larger than those along the fibers  ($\varepsilon_{11}< \varepsilon_{22}$), since $\varepsilon_{11}=P(1/E_1-\nu_{12}/E_{1})$ and $\varepsilon_{22}=P(1/E_2-\nu_{12}/E_{1})$. 
In planar structures, this behavior results in a pronounced out-of-plane deformation that can be controlled by tuning the orientation of the fibers \cite{Gladman_Lewis_2016}. 
Additionally, in cylindrical structures by varying the fiber angle a broad spectrum of deformations can be induced upon inflation, such as axial extension, radial expansion, and twisting~\cite{Suzumori_1991_fiber_reinforced, MICHALEK_Cohen_2019,  Connolly_Bertoldi_2017_fiber_reinforced,Connolly_Bertoldi_2015_Soft_Robot,Schaffner_NatComm2018, Singh_Girish_Soft_Robotics_2020, Wang_Polygerinos_IEEE_2017}.

\noindent {\it Knits. ---} 
Knitted fabrics are composed of fibers that are interwoven and entangled rather than embedded in a soft matrix (see Fig. \ref{fig1}b). They achieve flexibility through geometrically compliant interlooping stitch architectures, even when the yarns themselves are inextensible~\cite{Lechenault_PRX, Maziz_Sci_Adv_2017, Zulifqar_Textile_2019, Cappello_Walsh_Soft_Matter_2018, PENG2005859, Wicaksono_npj_Electron_2020, Phan_Do_Sci_Rep_2022, Liu_Textile_2010}. Additionally, these fabrics can be self-deforming if made using functional yarns \cite{Abel_Brei_SMS_2012, Abel_Brei_SMS_2013, Maziz_Sci_Adv_2017, Han_Hoon_Adv_Mater_2017}. Knitted fabrics are inherently anisotropic due to their looped structure and hold promise for creating shape-morphing metamaterials, but few practical examples have been reported to date~\cite{Sanchez_AdvFuncMat2023}.

\noindent {\it Tubes. ---} Thin elastic tubes made of isotropic elastic materials deform in an anisotropic manner upon inflation (see Fig.~\ref{fig1}c). For a thin and long inflated elastic tube with a radius $R$ and thickness $t$, the strains in the linear elastic regime along the circumferential and longitudinal directions are given by \cite{Timoshenko_book}:
\begin{equation}\label{strain_tubes}
\varepsilon_l = \frac{PR}{2tE}(1 - 2\nu), \quad\quad\quad \varepsilon_c = \frac{PR}{tE}(2 - \nu),
\end{equation}
where $P$ denotes the applied pressure and $E$ and $\nu$ are Young's modulus and Poisson's ratio of the material, respectively. Equation (\ref{strain_tubes}) illustrates that in inflated elastic tubes, the longitudinal strain is consistently smaller than the circumferential strain, and for an incompressible material ($\nu=0.5$), $\varepsilon_l$ vanishes. Importantly, arranging these tubes within a plane can induce out-of-plane deformation, which can be precisely controlled by adjusting the orientation of the tubes~\cite{Baromorphs, Siefert_Roman_Soft_Matter_2020, Gao_Bico_Roman_Science_2023, Panetta_ACMTransGraphics2021, Jones_PRL2023}.

\noindent {\it Anisotropic foams. ---} Anisotropic foams consist of a network of plates or beams that are preferentially oriented in one direction to impart directional properties (Fig.~\ref{fig1}d)~\cite{Lefebvre_foam_2017,Lefebvre_Martinez_foam_2016, van_Meerbeek_Shepherd_Adv_Mater_2016}. By designing foams with spatially varying mechanical properties, one can control how the foam deforms when subjected to external forces or stimuli~\cite{Lefebvre_ACMTranGraph2020,Vanneste_IEEERobot2020, van_Meerbeek_Shepherd_Adv_Mater_2016}. This ability opens up new possibilities for creating materials that can change shape in specific, controlled ways.

\subsection{Unit cells based on internal rotations} \label{section2.2}

%\url{https://onlinelibrary.wiley.com/doi/full/10.1002/advs.202204733}

Inspired by Ron Resch's early discoveries, which demonstrated the potential of using rotational motion to realize shape-morphing metamaterials \cite{RonResch_1973, Ron_Resch_patent1}, unit cells based on internal rotations typically consist of stiffer elements, such as rods or plates, connected by flexible hinges. These stiffer elements generally behave as nearly rigid, with deformation primarily localized at the hinges. Consequently, the shape change of these units can be described using relatively simple kinematic models that account for the connectivity and shape of the units. Nevertheless,  these hinges are implemented in practice using thin and flexible elastic components, whose elasticity can significantly influence the intended kinematics. Evaluating the effect of hinge elasticity on unit cell deformation involves simulating its behavior. To this aim, discrete models based on a combination of rotational and longitudinal springs to simulate the hinges' response have demonstrated high effectiveness~\cite{Bolei_PNAS_2020, Ishibashi_Iwata_2000, Coulais_NatPhys2018, Czajkowski_Coulais_2022, Corentin_Nature_2023, Filipov_Paulino_2015_origami_folding, FILIPOV_IJSS_2017_origami}.

\begin{figure}
	\centering
	\includegraphics[width=0.9\textwidth]{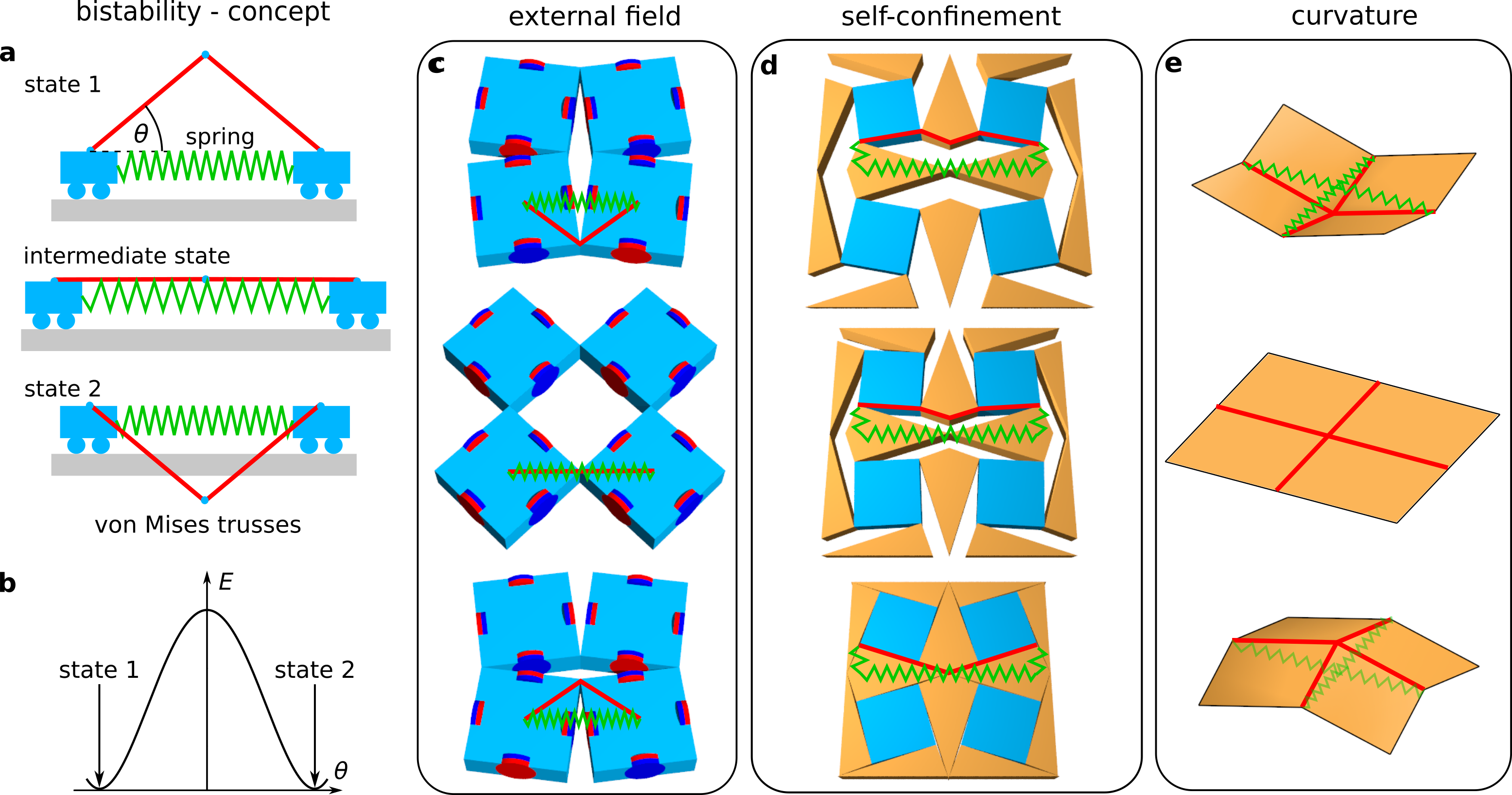}
	\caption{{\bf Bistable unit cells.}
		%Types of unit cells classified from the point of view of different means of achieving control over their configuration. 
		{\bf a} Bistable von Mises truss, made from two inextensible bars that hinge about sliding boundaries (represented by carts). The central spring has a rest length shorter than the added length of the trusses: this makes the flat configuration unstable and instead makes the popped up and popped down configurations stable. 
		{\bf b} Corresponding elastic energy vs. the angle $\theta$. All examples shown in panels {\bf c-e} can be mapped to such a von Mises truss.
		{\bf c} Magnetic bistable unit cell, where interactions between the rotating units are realized by magnetic interactions (adapted from \cite{Yasuda_Korpas_Raney_Phys_Rev_Appl_2020}). The magnets from adjacent squares create an attractive interaction, which makes the unfolded configuration unstable and the two folded configurations stable.
		{\bf d} Self-confining bistable unit cell, where the unfolding of the unit cell is not kinematically compatible and forces the rotating blocks to stretch (adapted from \cite{Ahmad_EML_2016}).  
		{\bf e} Bistable origami unit cell, where self-confinement arises because the vertex has a defect angle (the angles of the plates about the vertex add up to less than $2\pi$). As a result, the plates deform elastically when the vertex is forced through the flat configuration, which is then unstable. The popped-up and down configurations are then both stable, see \cite{Faber_Arrieta_Studart_Science_2018} for an example.
		On panels {\bf c-e}, auxiliary red lines are used to indicate representative von Mises trusses. 
	}
	\label{fig2}
\end{figure}

Unit cells utilizing internal rotations can exhibit either monostable or bistable behaviors. Monostable unit cells return to their initial configuration once the load is removed and thus require continuous actuation to maintain a deployed shape. In contrast, bistable unit cells possess multiple stable configurations, and in turn can be toggled between various stable shapes~\cite{Oppenheimer_PRE2015, David_Nature_2021, Meeussen_vanHecke_Nature_2023, Yasuda_Yang_PRL_2015, Liu_Paulino_Nat_Commun_2019}. 
Such multistable unit cells exploit the behavior of a von Mises truss---a structure comprising two rigid rods of length $L_{\rm rod}$  that are free to rotate with respect to each other at one end and are connected by a spring of stiffness $k$ and rest length $L_0$ at the other one (Fig. \ref{fig2}a,b). The  energy of such von Mises truss is given by
\begin{equation}
\mathcal{E}=\frac{k}{2}(L_0-2 L_{\rm rod} \cos\theta)^2
\end{equation}
and, as shown in Fig.~\ref{fig2}b, it exhibits two minima separated by an energy barrier. The minima corresponds to the two stable configurations with the truss pointing up or down and the maximum to the truss being flat. This suggests that one requires two minimal ingredients to achieve bistability for unit cells based on rotating unit cells: (i) a structure that can switch between two different configurations on either side of a flat state; (ii) some form of confinement. As we will see below, there are multiple ways of achieving such bistability (Fig.~\ref{fig2}c-e).

\begin{figure}
	\centering
	\includegraphics[width=0.98\textwidth]{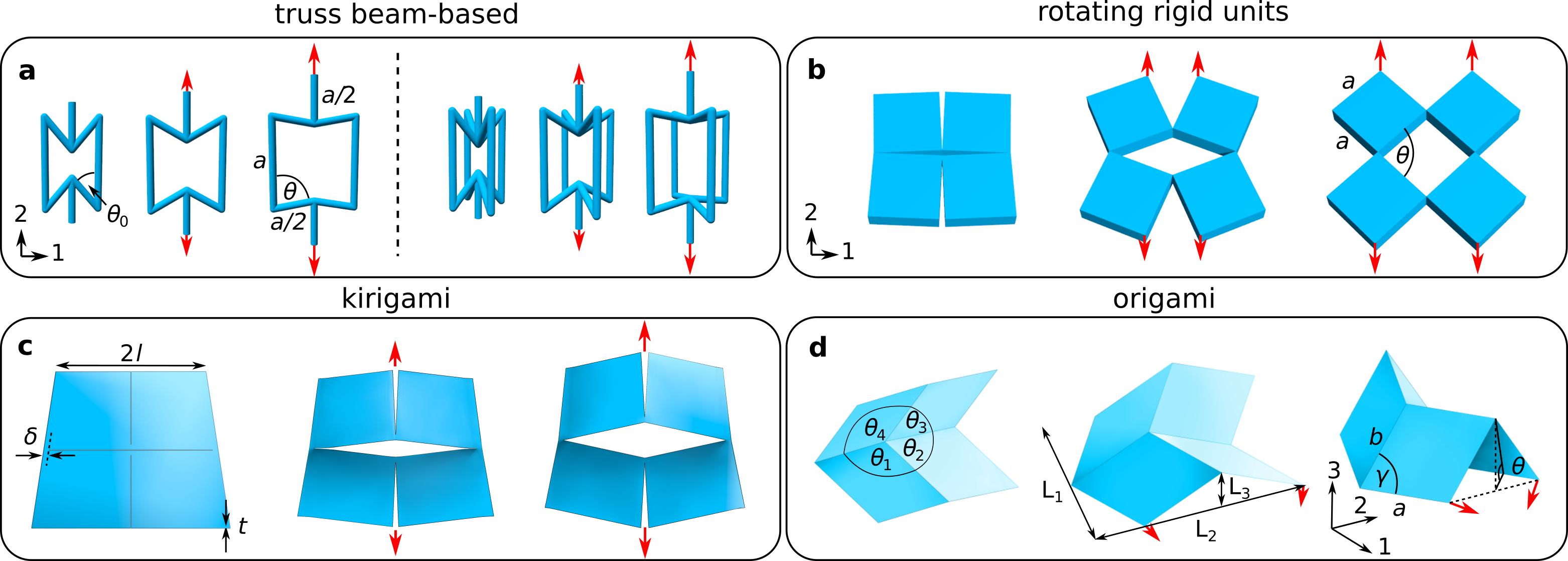}
	\caption{\textbf{Unit cells based on internal rotations.} Panels show unit cells made from {\bf a} rods, {\bf b} rotating blocks, {\bf c} kirigami and {\bf d} origami that can be used to assemble larger rotation-based shape morphing metamaterials.  Red arrows on all panels indicate the applied loading.
	}
	\label{fig3}
\end{figure}

{\it Rod-based unit cells} 
One of the best-known designs of rotating rod-based unit cells is that of the re-entrant hexagon, i.e. a unit cell composed of 8 rigid rods that can rotate relative to each other~\cite{Almgren1985}. Unlike a regular hexagon, which has all its angles pointing outward, a re-entrant hexagon has some angles that point inward, creating a non-convex shape (see Fig. \ref{fig3}a). In the ideal scenario of re-entrant hexagonal unit cells composed of rigid rods with lengths 
$a$ and $a/2$, which are free to rotate at their joints and initially form an angle $\theta_0$, the application of a displacement along the 
direction $x_2$ changes the angle between the rods to $\theta$. This results in the following expressions for the strains $\varepsilon_{11}$ and $\varepsilon_{22}$ \cite{MASTERS_Evans_1996}: %\KB{update the figure to reflect this notation. Sometimes we are using $\varepsilon_{xx}$ and sometimes $\varepsilon_{11}$. Let's pick one and be consistent}\KB{are the strains below correct?}
\begin{equation}
\varepsilon_{11} =  \frac{\sin\theta-\sin\theta_0}{\sin \theta_0},\;\;\;\; \varepsilon_{22} = \frac{ \cos \theta_{0} - \cos \theta}{2 - \cos \theta_{0}}.
\end{equation} 
It follows that the Poisson's ratio for loading in the direction $x_2$ is given by
\begin{equation}
\nu_{21} =-\frac{\varepsilon_{11}}{\varepsilon_{22}}= - \frac{(\sin \theta - \sin \theta_{0})(2  - \cos \theta_0)}{(\cos\theta_0 - \cos \theta)\sin \theta_{0}}
\end{equation}
which is negative for any value of $\theta$ and $\theta_{0}$ in the interval between $0^{\circ}$ and $90^{\circ}$. While most materials and structures exhibit a positive Poisson’s ratio, systems with a negative Poisson’s ratio (known as auxetic) behave differently: they contract (expand) in the transverse direction when compressed (stretched). Notably, auxeticity not only enables unusual shape changes within the plane of the structures but can also be leveraged to control their Gaussian curvature \cite{Mirzaali_Zadpoor_Adv_Mater_2021, Wei_Mahadevan_PRL_2013}. 

The re-entrant honeycomb structure is not the only rod-based unit cell that has been suggested. Numerous other rod-based unit cells with a similar hinging-dominated deformation mechanism have also been proposed \cite{Larsen_Sigmund_1997, YANG_Cormier_IJSS_2015, Hengsbach_SMS_2014,Lim_arrowhead_2016,Grima_star_2005_Mol_Simul,Grima_egg_rack_2005, Wu_Berto_Mater_Des_2019}. The shape morphing capabilities of rod-based unit cells can be further enhanced by making them bistable and able to undergo rapid snap-through buckling~\cite{Jin_Slocum_bistable_beam_2004, Rafsanjani_Pasini_Adv_Mater}.

{\it Units cell based on rotating blocks } %\KB{the text here is rough, but these are the points that I think we can make. Arethere otehr points that you would like to add?}
Rigid units connected at their vertices through hinges can rotate with respect to each other to produce large changes in their covered area.  The most famous example of such unit cells is that of the so-called rotating squares \cite{Grima_Evans_squares_2000, Bertoldi_Mullin_Adv_Mater_2010_instabilities_Poisson}, which consist of four rigid squares connected at their vertices through perfect hinges (see Fig. \ref{fig3}b).  When constrained to move in a plane, linear dimensions of such unit cells are given by 
\begin{equation}
L_{1}=L_2 = 2a \left[\cos \left(\frac{\theta}{2} \right) + \sin \left(\frac{\theta}{2} \right) \right]
\end{equation}
where $a$ denotes the edge length of the squares and $\theta$ is the angle between the squares. It follows that this ideal unit cell exhibits a large negative Poisson's ratio that irrespective of the loading direction is given by 
\begin{equation}
\nu=-\frac{\varepsilon_{11}}{\varepsilon_{22}}=-1.
\end{equation}
Numerous strategies have been proposed to realize this mechanism both in its open (i.e., with 
$\theta\gg 0$) and closed (i.e., with 
$\theta\approx 0$) configuration. The open configuration can be fabricated by embedding arrays of holes in elastic matrices \cite{Bertoldi_Mullin_Adv_Mater_2010_instabilities_Poisson, Mullin_Bertoldi_Boyce_Phys_Rev_Lett_2007} and connecting rigid units through flexible components \cite{Deng_Bertoldi_Nat_Commun_2018}. The closed configuration can be manufactured by embedding arrays of cuts into elastic sheets \cite{Rafsanjani_Bertoldi_kirigami_PRL_2017}. In the former, compression is applied to achieve auxetic behavior, typically triggered by instability \cite{Bertoldi_Mullin_Adv_Mater_2010_instabilities_Poisson}, while in the latter, auxeticity occurs under tensile loading. Regardless of the fabrication method, none of the unit cells achieve the ideal Poisson's ratio of -1, as the elasticity of the hinges limits their rotations. 

In addition to the rotating squares, a variety of unit cells based on hinged rigid  2D polygons~\cite{Grima_Evans_pssb_2005, Grima_Evans_triangles_2006, MILTON20131543, Singh_van_Hecke_PRL_2021, Bossart_PNAS_2021, Grant_website} and 3D polyhedra~\cite{Ron_Resch_movie, Kim_PRSA_2017_3D_aux_prisims} 
have been proposed. However, for these more general structures, Poisson’s ratios are typically not constant and depend both on the direction of loading and the amount of applied deformation.

While unit cells based on rotating blocks are usually monostable, they can also be made bistable \cite{Deng_Bertoldi_Sci_Adv_2020,Yasuda_Korpas_Raney_Phys_Rev_Appl_2020, Peng_Plucinsky_Phys_Rev_Appl_2024}. In particular, bistable designs of rotating squares structures have been realized by adding elastic springs ~\cite{Rabago_Overvelde_EML_2022_bistable_hinges, Deng_Bertoldi_Sci_Adv_2020} or permanent magnets (Fig.~\ref{fig2}c)~\cite{Yasuda_Korpas_Raney_Phys_Rev_Appl_2020,Dudek_Gatt_3D_Mater_Des_2020, Korpas_Yasuda_Raney_ACS_Appl_Mater_Interf_2021}, to give an energy cost to the lateral expansion of pairs of squares as $\theta$ decreases, effectively forming a von Mises truss.  Furthermore, it has been demonstrated that the confinement provided by the surrounding elements (Fig.~\ref{fig2}d) can effectively lead to the formation of bistable von Mises-like truss mechanisms within the unit cells \cite{RAFSANJANI_Pasini_EML_2016, Singh_van_Hecke_PRL_2021}.

{\it Kirigami unit cells} Kirigami unit cells are created by introducing arrays of cuts into elastic sheets, effectively forming hinges where rotations localize, ultimately leading to significant changes in shape \cite{Blees_McEuen_Nature_2015, Chen_Wu_Mater_Des_2019, Edini_Paulino_Sci_Adv_2015, Neville_Scarpa_Sci_Rep_2016}. When the elastic sheet is thick, kirigami unit cells deform planarly and are equivalent to unit cells based on rotating units realized in their closed configuration. However, as the thickness $t$ of the sheet decreases, tensile loading may induce out-of-plane buckling of the hinges, leading to the formation of complex 3D patterns \cite{Tang_Kamien_Yin_Adv_Mater_2017, Xu_Lopez_Adv_Mater_2021, Rafsanjani_Bertoldi_kirigami_PRL_2017} at a critical strain \cite{Isobe_Okumura_Sci_Rep_2016, Rafsanjani_Bertoldi_kirigami_PRL_2017}
\begin{equation}
\varepsilon_{cr} \propto  \left(\frac{t}{\delta} \right)^{2}
\end{equation}
where $\delta$ represents the width of the hinges across the plane. 
The simplest thin kirigami unit cell consists of a series of parallel cuts \cite{Lamoureux_Nat_Commun_2015}. However, numerous other patterns have been proposed to achieve complex reconfigurations \cite{Xu_Lopez_Adv_Mater_2021, Cho_Srolovitz_PNAS_2014_hierarchical, Tang_Yin_EML_2017, Liu_Du_Sci_Adv_2018}, including that of the rotating squares  (see Fig. \ref{fig3}c) \cite{Rafsanjani_Bertoldi_kirigami_PRL_2017, Tang_Li_PNAS_2019}.

Finally, kirigami unit cells can be made multistable \cite{Yi_Dias_Holmes_PhysRevMater_2018, Lele_Mayers_Comp_Sci_Technol_2019, Yang_Chen_APL_2023_multistable_kirigami}. Most notably, for the standard kirigami pattern created via a series of parallel cuts \cite{Yi_Dias_Holmes_PhysRevMater_2018}, the bistable behavior of unit cells originates from the possibility of adjacent square-like panels to bend out-of-plane to assume stable symmetric or antisymmetric configurations. However, kirigami created by parallel cuts can also adopt more complex configurations \cite{Lele_Mayers_Comp_Sci_Technol_2019, Yang_Chen_APL_2023_multistable_kirigami,Janbaz_Coulais_Nat_Commun_2024}. In such scenarios, it is possible to observe even four stable configurations as a result of the appropriately distributed parallel cuts.

{\it Origami unit cells} 
In rigid foldable origami, the panels (or facets) between the creases remain rigid and do not deform. Folding and unfolding occur solely through rotation around the crease lines (hinges), which act as pivot points. In contrast, folding of a non-rigid foldable origami (also known as flexible or compliant origami) involves not only the rotation of the panels around the crease lines but also their bending and stretching. The Miura-ori~\cite{Miura_1985} and the waterbomb~\cite{Hanna_Howell_SMS_2014} patterns are among the most well-known examples of rigid foldable origami metamaterials. In contrast, non-rigid foldable origami structures include square-twist origami \cite{Silverberg_Cohen_Nat_Mater_2015, Ma_Zhong_IJMS_2021, Zang_Chen_Thin_Walled_Struct_2022} and Kresling kirigami \cite{Cai_Zhou_J_Mech_Robot_2017, Berre_Renaud_J_Mech_Robot_2022}. However, with specific geometric parameters, these non-rigid designs can also be engineered to exhibit rigid foldability \cite{Pagano_Tawfick_SMS_2017}. 

For rigid origami, the configurations induced by folding can be determined by simply applying trigonometry. For example, for the Miura-ori pattern \cite{Wei_Mahadevan_PRL_2013, Schenk_Guest_PNAS_2013}, the linear dimensions of the unit cell  can be expressed as (see Fig. \ref{fig3}d):

\begin{equation*}
L_{1} = \frac{2b \cos(\theta) \tan(\gamma)}{\sqrt{1 + \cos^{2}(\theta) + \tan^{2}(\gamma) }} ; \;\;\; L_{2} = 2a \sqrt{1 - \sin^{2}(\theta) \sin^{2}(\gamma)} + \frac{b}{\sqrt{1 + \cos^{2}(\theta) \tan^{2}(\gamma)}} ;\;\;\; L_{3} = a \sin(\theta) \sin(\gamma).
\end{equation*}

\noindent
where, $a$, $b$, and $\gamma$ are side lengths and the internal acute angle of the rigid facets. Furthermore, $\theta\in [0, 90^{\circ}]$ is the angle between the facets and the $x_1-x_2$ plane. On the other hand, for the non-rigid foldable origami, the deformation process is more complex and typically requires the use of the nonlinear model to describe it \cite{Liu_Paulino_PRSA_2017}.
%\KB{how about non-rigid origami patterns?}

Another important classification of origami unit cells is the distinction between flat-foldable and non-flat-foldable designs \cite{Misseroni_Paulinho_Nat_Rev_2024}. Flat-foldable origami, such as the Miura-ori pattern  \cite{Miura_1985, Schenk_Guest_PNAS_2013}, can be completely folded into a flat, two-dimensional shape. This characteristic is particularly useful in applications where compact storage or transport is necessary, and the structure needs to be deployed or expanded when in use. In contrast, non-flat foldable origami, such as the axisymmetric variant of the Miura origami \cite{Dang_Paulino_PRSA_2024}, does not fold completely flat but instead transforms into three-dimensional shapes with various functionalities. Researchers have attempted to determine the conditions under which flat-foldability of an origami unit cell can be achieved \cite{Huffman_1976}. Generally, this task is very challenging, but such conditions can be determined for some specific origami patterns such as the vertex-4 pattern \cite{Hull_2003_flat_foldability, Evans_Howell_Royal_Soc_2015, Schenk_Guest_PNAS_2013}. 

%\cc{ADD A NOTE ABOUT BISTABLE VANILLA ORIGAMI: SQUARE TWIST, MIURA-ORI.}
Origami unit cells can also exhibit bistable behavior. Similar to rotating rigid block metamaterials, bistable rigid foldable origami can be achieved by exploiting magnetic interactions between strategically distributed magnetic inclusions \cite{Fang_Wang_SMS_2020}. More intriguingly, the deformation of faces and hinges that occurs in non-rigid foldable origami can be harnessed to achieve bistability. In non-rigid foldable origami, the faces typically bend to accommodate the folding motion~\cite{Silverberg_Cohen_Science_2014, Silverberg_Cohen_Nat_Mater_2015}.
As a result of face bending in these cases, the structures form von Mises trusses, where they may need to overcome an energy maximum (the elastic energy from face bending) and can thus become bistable. Alternatively, the stretching and shearing of the hinges can be exploited to design origami with two stable states separated by an energy barrier~\cite{David_Nature_2021}.
Another approach is non-Euclidean origami~\cite{Waitukaitis_van_Hecke_PRE_2020, Santangelo_Annual_Review_2017}, where the sector angles around each vertex sum to $2\pi+\varepsilon$ instead of $2\pi$. An excess angle $\varepsilon > 0$ results in a saddle shape, while a deficit angle $\varepsilon < 0$ leads to a cone shape (see Fig. \ref{fig2}e). This non-Euclidean property makes the flat state energetically unstable, enabling multiple folded stable configurations. This approach has been used to achieve bistable~\cite{Hanna_Howell_SMS_2014, Silverberg_Cohen_Nat_Mater_2015, Fang_Wang_PRE_2017, Faber_Arrieta_Studart_Science_2018, Rabago_Overvelde_Nat_Commun_2019} and even tristable behavior~\cite{Waitukaitis_van_Hecke_PRE_2020}.

\section{Different strategies of assembling unit-cells into shape-morphing metamaterials}

Mechanical metamaterials capable of shape-morphing are designed by tessellating the unit cells described in Section \ref{unitcells}. To achieve precise, on-demand shape changes, both the geometry of the individual unit cells, as well as their spatial arrangement, are critical factors. As we will see, shape morphing can be either monostable—requiring continuous actuation to maintain the shape—or multistable, where the shape is retained even after actuation is turned off. This behavior depends not only on the intrinsic multistability of the unit cells but can also emerge from the collective interaction between them.

To provide a unified framework for designing assemblies of unit cells, we propose a classification based on two criteria: ($i$) whether the deformations of neighboring unit cells are kinematically compatible or incompatible (i.e., whether they are geometrically compatible or frustrated), and ($ii$) whether the assembly consists of a single periodically tiled unit cell or a combination of various unit cells in the tiling.

\subsection{Mechanical metamaterial consisting of compatible unit-cells}

We begin by examining metamaterials composed of unit cells that exhibit collective deformations, where each unit cell deforms according to its low-energy mode in a kinematically compatible manner. In this scenario, as previously mentioned, there are two possible sub-cases: periodic tessellations and non-periodic tessellations of the unit cells.

\subsubsection{Periodic tesselations}

The simplest and most widespread way to create a metamaterial is to periodically tile space with a unit cell. For example, take any of the unit cells in Fig.~\ref{fig1} and Fig.~\ref{fig2}, tile them in a 2D or 3D lattice and you obtain a metamaterial that will deform following the deformations of the unit cells if the loading is homogeneous. Those deformations are typically large as a result of the compliance of the unit cells and can lead to either pronounced shear deformations or volume changes. However, irrespective of the type of reconfiguration, these deformations are homogeneous at the level of the entire metamaterial and therefore lead to limited shape-changing capabilities. With periodic tilings, one can achieve complex on-demand shape-change by exploiting controlled spatial variations of the strain field via (i) local instabilities, (ii) boundary conditions, or (iii) defects. We briefly discuss these strategies below.

\begin{figure}[t!]
	\centering
	\includegraphics[width=0.98\textwidth]{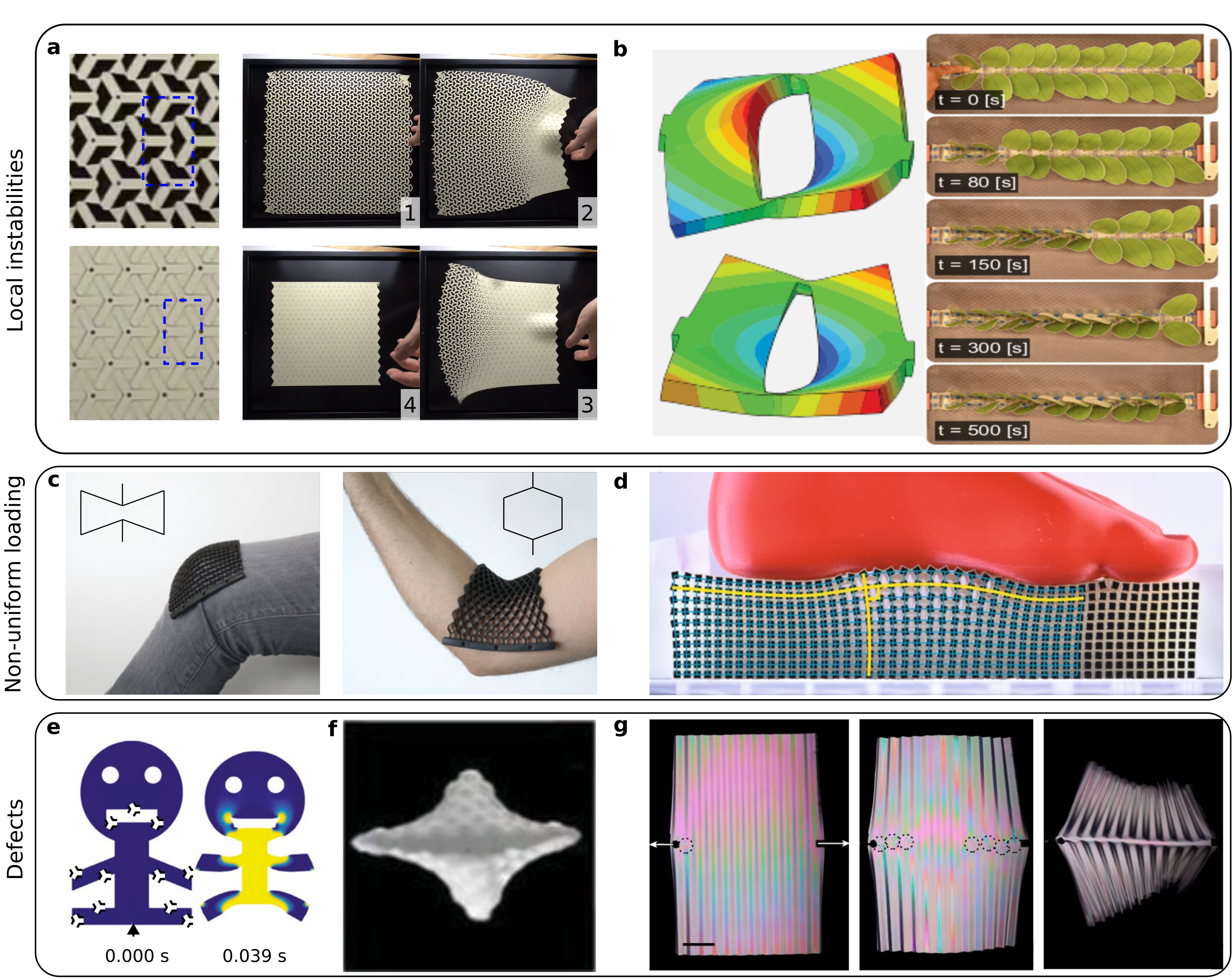}
	\caption{\textbf{Kinematically compatible metamaterials made from periodic tesselations.} {\bf a-b} Examples of periodic tessellations that achieve complex on-demand shape-change by exploiting local instabilities. {\bf a} In-plane kirigami with a bistable unit cell exhibits an inertial transition wave that takes it from an open to a compact state \cite{Lishuai_Kochmann_PNAS_2020}. {\bf b} Kirigami metamaterial composed of bistable unit cells that sequentially undergo a diffusive transition from the antisymmetric to the symmetric configuration \cite{Janbaz_Coulais_Nat_Commun_2024}. The kirigami is dressed with leaves and mimics the sequential diffusive folding of the mimosa pudica plant. {\bf c-d} Examples of periodic tessellations that achieve  complex on-demand shape-change by exploiting boundary conditions. {\bf c} Metamaterials drape surfaces with positive or negative gaussian curvature depending on their Poisson's ratio \cite{Mirzaali_Zadpoor_Adv_Mater_2021}. {\bf d} Auxetic metamaterials under non-homogeneous loading accomodate conformal deformations (no shear, only dilation and gradients thereof) \cite{Czajkowski_Coulais_2022}. {\bf e-g} Examples of periodic tessellations that achieve complex on-demand shape-change by exploiting defects that create a strain field around them. {\bf e} Strategic placement of defects enables control over the shape changes \cite{Lishuai_Kochmann_PNAS_2020}. {\bf f} Popping domes in dimpled sheets create complex plate bending \cite{Faber_Arrieta_Adv_Sci_2020_shells}. {\bf g} Pulling on a corrugated sheet creates collections of mobile defects that lead to a zoo of shapes \cite{Meeussen_vanHecke_Nature_2023}.}
	\label{fig4_periodic}
\end{figure}

\emph{Local instabilities. ---}
The softening caused by snap-through buckling in the unit cells generates metamaterials that deform through a sequence of events~\cite{Rafsanjani_Pasini_Adv_Mater, Ahmad_PNAS_2020} (Fig. \ref{fig4_periodic}a) and can be utilized for gradual shape-change applications in soft robotics (Fig.\ref{fig4_periodic}b)\cite{Rafsanjani_Bertoldi_PNAS_2019, Janbaz_Coulais_Nat_Commun_2024}. While most metamaterials exhibiting sequential deformation are constructed using multistable unit cells, it is crucial to understand that multistability is not essential for achieving sequential behavior. The critical factor is softening, which can occur through snap-through buckling even in a monostable structure~\cite{Rafsanjani_Bertoldi_PNAS_2019}.

Furthermore, sequential deformation can also arise in hierarchical metamaterials composed of slender elements with varying critical loads. In these systems, different components buckle in sequence once their critical Euler load is exceeded, leading to a series of shape transformations~\cite{Coulais_Sabbadini_Nature_2018}.

\emph{Boundary conditions---} Complex on-demand shape-changes can also be triggered by carefully selecting the boundary conditions on the metamaterial. Just as a piece of ordinary hyperelastic material, a metamaterial can take arbitrarily complex shapes when a complex load distribution is applied to it. The main difference with ordinary materials though is that the deformations are large as a result of the high compliance of the metamaterial and that specific modes of deformations are favored that differ from that of ordinary materials. Take for instance the simple bending of a slender object, when bent downwards, a plate of rubber will tend to warp upwards on the side as a result of its incompressibility, thus displaying negative Gaussian curvature. In contrast, an auxetic metamaterial will warp downwards because of its auxetic nature, thus displaying positive Gaussian curvature~\cite{Wei_Mahadevan_PRL_2013,Mirzaali_Zadpoor_Adv_Mater_2021}. In fact, as a result of their enhanced compressibility, auxetic metamaterials can more readily change their Gaussian curvature and drape complex surfaces with positive Gaussian curvature, whereas non-auxetic metamaterials can drape complex surfaces with negative Gaussian curvature (Fig. \ref{fig4_periodic}c).
In addition, deformations of auxetic metamaterials typically consist of dilations, whereas gradients of dilation and shear deformations can be neglected. As a result, theoretical tools such as conformal maps can be used to predict and reverse their shape-changes~\cite{Czajkowski_Coulais_2022} (Fig. \ref{fig4_periodic}d). Hence, especially metamaterials composed of rotation-based unit cells are very sensitive to the application of the nonuniform force and can undergo complex conformal transformations \cite{Cho_Srolovitz_PNAS_2014_hierarchical, Hwang_Bartlett_2022}. More generally, metamaterials can be designed to exhibit dominant modes of deformation with enhanced compliance, deformation range, and anisotropy, therefore they can exhibit a wide palette of large and target deformations under complex loading.

\emph{Defects that create a strain field. ---} A third approach involves strategically introducing defects into rotation-based unit cells. These defects create a strain gradient around them, which can be used to induce either in-plane~\cite{Silverberg_Cohen_Science_2014, Lishuai_Kochmann_PNAS_2020} or out-of-plane~\cite{Faber_Arrieta_Adv_Sci_2020_shells, M_Liu_2023_JMPS_snap_induced_morphing,Meeussen_vanHecke_Nature_2023} shape changes. Note that the defects can be rigid objects that are initially seeded within the structure~\cite{Lishuai_Kochmann_PNAS_2020} (Fig. \ref{fig4_periodic}e), or they can be bistable unit cells that can be "popped through" on demand~\cite{Silverberg_Cohen_Science_2014, Faber_Arrieta_Adv_Sci_2020_shells, M_Liu_2023_JMPS_snap_induced_morphing} (Fig. \ref{fig4_periodic}f). Additionally, defects can even spontaneously emerge when a simple uniaxial load is applied to the metamaterial~\cite{Meeussen_vanHecke_Nature_2023} (Fig. \ref{fig4_periodic}g). In this latter case, the defects can move and nucleate in sequence, enabling a wide variety of shapes to be achieved with simple loading.

However, it is important to note that predicting the strain gradient in the regime of large deformations is extremely challenging, making the inverse design of these metamaterials particularly difficult.

\subsubsection{Non-periodic tesselations}

As previously discussed, periodic tilings can result in complex shape morphing, but predicting their behavior is challenging due to nonlinear effects such as instabilities. To better design shape-morphing metamaterials, one approach is to assemble dissimilar unit cells. The idea that combining structural components with different mechanical responses can produce interesting shape changes has been recognized for a long time. A classic example is the bimetallic strip~\cite{Timoshenko_1925_bimaterial}, which consists of two metal strips that expand at different rates when heated. When joined, these differing expansions cause the flat strip to bend in one direction when heated and in the opposite direction when cooled. This concept can be extended beyond slender structures to the realm of metamaterials. There are several ways to implement this idea. One approach is to spatially grade the unit cell geometry, such as by gradually changing the lattice spacing or unit cell geometry across the material. Another approach is to eliminate periodicity entirely and design fully disordered architectures. A third approach is to maintain some degree of order by designing lattices with consistent periodicity but combining unit cells with different topologies or orientations in a way that creates an aperiodic or quasi-crystalline structure. These three approaches are discussed below.

\begin{figure}[t!]
	\centering
	\includegraphics[width=0.99\textwidth]{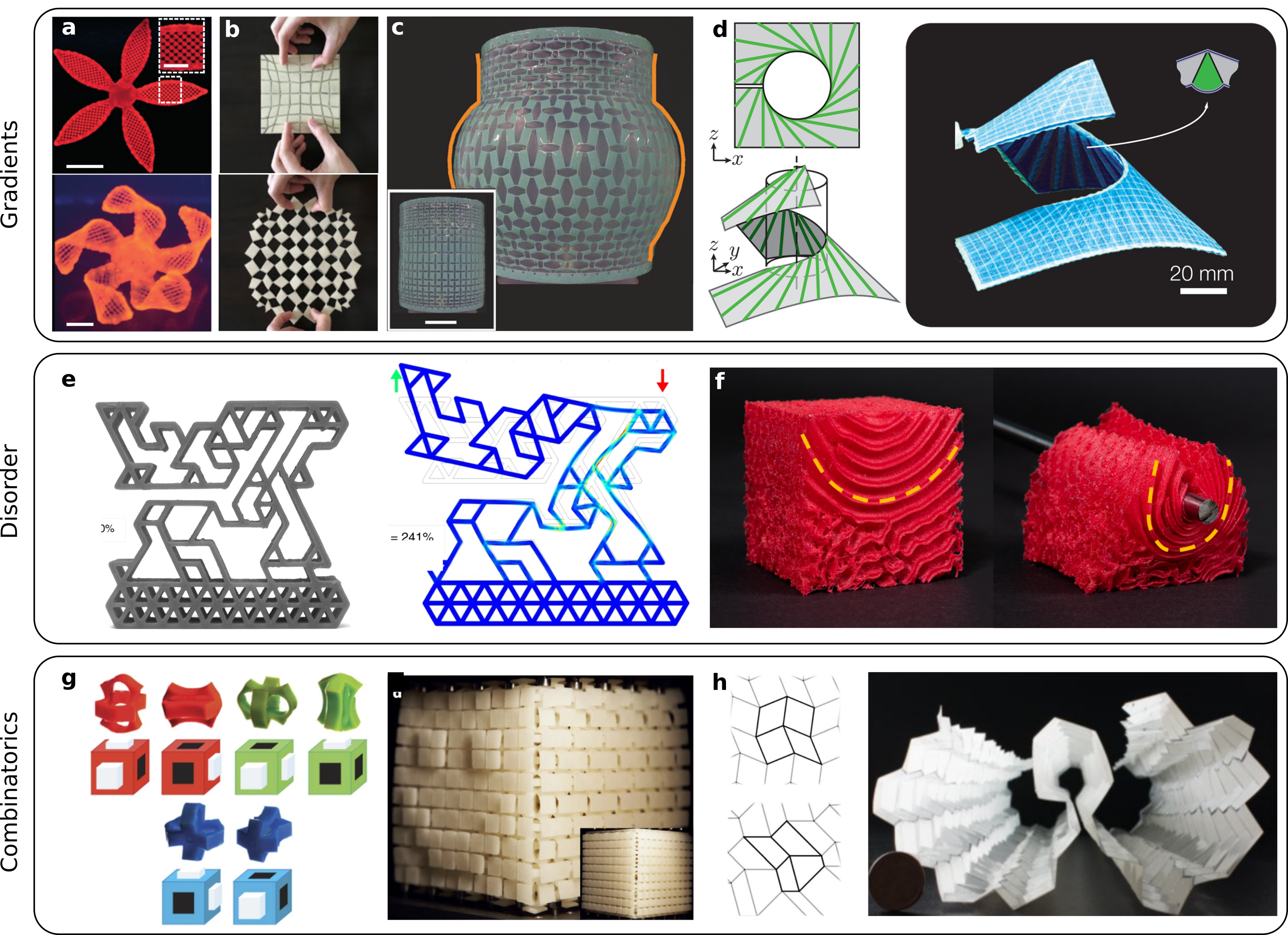}
	
	\caption{\textbf{Kinematically compatible metamaterials made from non-periodic tessellations.} 
		{\bf a-d}. Examples of non-periodic
		tessellations that achieve complex on-demand shape-change by exploiting gradients.
		{\bf a} Composite hydrogel architecture encoded with localized, anisotropic swelling behaviour controlled by the alignment of cellulose fibrils along prescribed pathways~\cite{Gladman_Lewis_2016}. 
		{\bf b} Kirigami pattern programmed for getting a square to a circle on deployment~\cite{Choi_Dudte_Nat_mater_2019}.
		{\bf c} Kirigami inflatable  that mimic target jar profiles  upon pneumatic actuation \cite{Lishuai_Bertoldi_2020OInflatable_kirigami}).
		{\bf d}  Developpable helicoid made from  trapezoidal channels 3D printed on a layer of airtight fabric ~\cite{Gao_Bico_Roman_Science_2023}.
		{\bf e-f} Examples of non-periodic
		tessellations that achieve complex on-demand shape-change by exploiting disorder.
		{\bf e} Disordered lattice programmed to achieve a target movement upon mechanical loading \cite{Bonfanti_Zapperi_Nat_Commun_2020}.
		{\bf f} Disordered  foam with fiber oriented to control how the  volume reshapes under
		large deformations \cite{Lefebvre_ACMTranGraph2020}. 
		{\bf g-h} Examples of non-periodic
		tessellations that achieve complex on-demand shape-change by exploiting combinatorics. 
		{\bf g} 3D cubic metamaterial made from aperiodically oriented building blocks reveals its precisely designed surface texture under uniaxial compression~\cite{Coulais_van_Hecke_Nature_2016}). 
		{\bf h} Combinatorial origami made from a discrete set of origami vertices can deploy into multiple target shapes~\cite{Dieleman_van_Hecke_2019}.}
	\label{fig5}
\end{figure}

\emph{Gradients. ---} A tessellation of spatially varying unit cells is expected to produce spatially varying shape changes, which collectively generate non-homogeneous strains and result in overall shape transformations at the structural level. For instance, controlling the arrangement of fibers within a soft matrix has been shown to achieve complex bending and twisting deformations in a thin sheet (Fig. \ref{fig5}a)~\cite{Gladman_Lewis_2016, Studart_NatComm2013}. However, this concept is not limited to slender structures and can also be applied to systems deforming in-plane.
For example, different distortions of the rotating square mechanism from Fig. \ref{fig1}b can be combined, ensuring that each unit cell still fully deploys~\cite{Choi_Dudte_Nat_mater_2019} (Fig. \ref{fig5}b). This enables varying deformations across the metamaterial, allowing for more pronounced deformation inside the structure than outside, which can be used to design a metamaterial capable of deploying from a square to a disk or adopting a specific 3D shape~\cite{Choi_Dudte_Nat_mater_2019}. This approach can be applied to a variety of rotation-based unit cells, including origami~\cite{Dudte_Vouga_2016}, kirigami~\cite{Lishuai_Bertoldi_2020OInflatable_kirigami,Tani_Roman_EML_2024}, cellular metamaterials~\cite{Dudek_Kadic_Adv_Mater_2022,Zhang_Krushynska_APL_2022, Mirzaali_Zadpoor_Sci_Rep_2018}, and anisotropic unit cells, such as inflatable tubes~\cite{Gao_Bico_Roman_Science_2023} (Fig. \ref{fig5}c).

\emph{Disorder. ---} Metamaterials with disordered lattice structures, lacking positional order, can also exhibit shape-morphing capabilities~\cite{Zapperi_review_2023, Grima_Mizzi_cuts_Adv_Mater_2016}. One example is disordered networks engineered to achieve a specific input/output relationship, a phenomenon known as allostery~\cite{Rocks_Pashine_PNAS_2017, Reid_Pashine_PNAS_2018, Bonfanti_Zapperi_Nat_Commun_2020} (Fig. \ref{fig5}e). Another example is anisotropic foams, composed of random microstructures without a repetitive unit cell (Fig. \ref{fig5}f). These foams can be designed using computer graphics techniques such as procedural textures to display an effective elastic tensor with a homogenized anisotropic response, enabling targeted shape changes~\cite{Lefebvre_ACMTranGraph2020}. 

\emph{Combinatorics. ---} A third approach involves arranging unit cells with different types or orientations on a periodic lattice to achieve controlled texture morphing~\cite{Coulais_Teomy_Nature_2016, An_Bertoldi_hierarchical_2020} (Fig. \ref{fig5}g). As there is only a finite number of possible configurations, such problems have an inherently discrete design space, in contrast with the aforementioned classes that had continuous design spaces. In addition, minute changes in the type of orientation of a single unit cell can completely change the response, e.g. make it switch from floppy to rigid~\cite{Coulais_Teomy_Nature_2016,Meeussen_NatPhys2020, Bossart_PNAS_2021,vanMastrigt_PRL2022} or change the number of low energy deformation modes~\cite{Dieleman_van_Hecke_2019, Bossart_PNAS_2021, vanMastrigt_PRL2022, vanmastrigt2024prospectingpluripotencymetamaterialdesign} (Fig. \ref{fig5}h). This sensitivity to minute changes renders the problem combinatorial in nature, where only a tiny subset of the design space leads to successful configurations with an on-demand shape-change. The interplay between shape-change and combinatorics can find analogues in other fields of science such as protein folding~\cite{huang2016coming, lovelock2022road}, self-assembly~\cite{zeravcic2014size, evans2024pattern}, computer graphics~\cite{merrell2007example, merrell2010model}, and molecular design~\cite{bilodeau2022generative}.

%\cite{}

\subsection{Mechanical metamaterials with kinematic incompatibility between unit cells}

The range of shapes that mechanical metamaterials can achieve can be expanded by arranging unit cells in ways that introduce geometric frustration into their interactions. This kinematic incompatibility between unit cells prevents the system from minimizing all elastic interactions simultaneously, often resulting in mechanical instabilities that can induce significant shape changes from small inputs. Since shape morphing in periodic tessellations of kinematically incompatible unit cells is primarily driven by instabilities, designing systems that achieve target shapes becomes more complex and requires a finer understanding of the nonlinear mechanics at play and more advanced optimization tools. Consequently, this approach has been less explored, but it holds intriguing potential: small inputs can trigger large shape changes, and multiple degenerate states can emerge, between which the metamaterial can switch.

%To conclude, both periodic and non-periodic tesselations with kinematic incompatibility buckle out of the plane to exhibit rapid shape changes or exhibit multistability. These two attributes offer exciting prospects for shape-morphing metamaterials as they allow either large shape changes with a small stimuli or the ability to retain shape even after unloading. As a result of their nonlinear and multistable response, these metamaterials are however much more difficult to design than their compatible counterpart and require a finer understanding of the nonlinear mechanics at play and more advanced optimization tools.

For metamaterials with kinematic incompatibility, as with compatible metamaterials, we classify them into two categories: those with unit cells organized through periodic tessellations and those organized through non-periodic tessellations.

\begin{figure}[t!]
	\centering
	\includegraphics[width=0.98\textwidth]{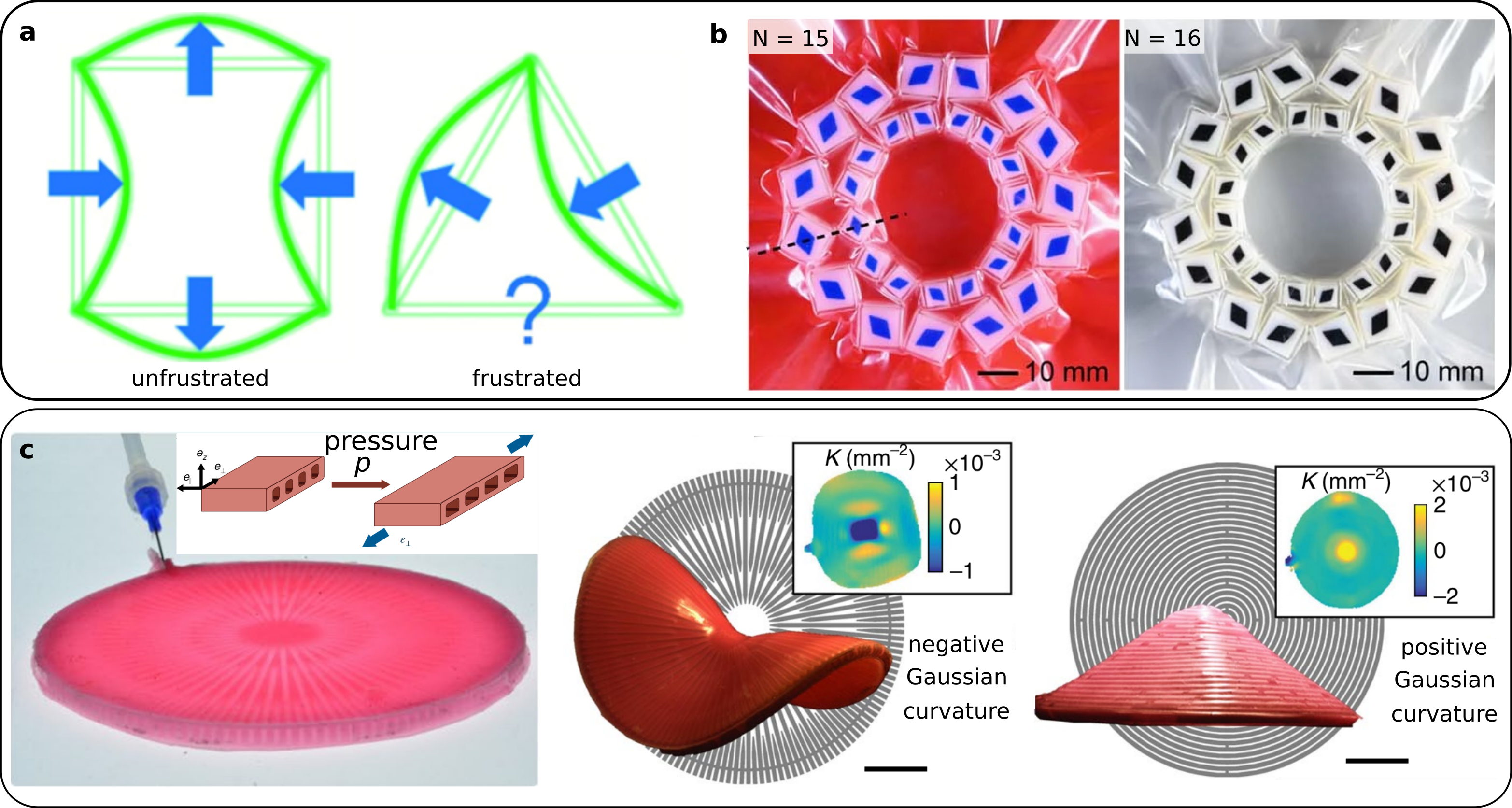}
	\caption{\textbf{Kinematically incompatible metamaterials made from periodic tessellations.} 
		{\bf a-b} Geometric frustration from loops with an odd number of units. {\bf a} Buckled beams on frames want to preserve angles at joints to minimize the deformation energy. This can be realized for square frames, but not for triangular frames which become geometrically frustrated~\cite{Kang_Clarke_Bertoldi_PRL_2014_instability}. {\bf b} This concept can be generalized to metamaterials made from global loops, where the parity of the loop defines whether the metamaterial is frustrated or not~\cite{Corentin_Nature_2023}. {\bf c} Frustrated plate-shaped metamaterial made of anisotropic tubes~\cite{Baromorphs}.}
	\label{fig6}
\end{figure}

\subsubsection{Periodic tesselations}
Geometric frustration in periodic tessellations of unit cells can be introduced by leveraging buckling in loops that impose a global constraint. This phenomenon can be illustrated by examining elastic beams connected to form two-dimensional square and triangular frames~\cite{Kang_Clarke_Bertoldi_PRL_2014_instability}. In the unfrustrated square frame, each beam can buckle into its lowest-energy configuration—a half sinusoid—while simultaneously preserving joint angles with neighboring beams, minimizing deformation energy (Fig. 6a). In a triangular frame, however, such configurations are unattainable, resulting in a frustrated system. Notably, it has been shown that buckling in frustrated triangular cellular structures leads to the formation of complex, ordered patterns~\cite{Kang_Clarke_Bertoldi_PRL_2014_instability}. 
%\KB{how about adding a small panel to Fig 6 to explain this?} \cc{sounds good to me! I guess one example is enough. (if we wanted another one we could add the example of buckling of a moebius strips, but that perhaps that would be too much?)}

\emph{Tessellations of  unit cells based on internal rotations. ---} 
%\KB{we should include an example in Fig 6} \cc{We could add Fig S2a of the paper "Non-orientable order and non-commutative response in frustrated metamaterials " to illustrate this.}
Geometric frustration can be introduced in periodic tessellations of rotating squares by creating loops with an odd number of units. In a structure made of rotating squares, the energy is minimized when neighboring units rotate in opposite directions. However, this preferred deformation mode cannot be supported in cylinders \cite{Javid_Bertoldi_JMPS_20161, Zhou_Chen_Commun_Mater_2024}, rings, or tori~\cite{Corentin_Nature_2023} with loops containing an odd number of squares. Along these loops, local constraints cannot all be satisfied simultaneously, as not all neighboring square pairs can rotate in opposite direction. This incompatibility between the optimal deformation mode of the rotating squares and the odd number of units in the loops makes the system geometrically frustrated, resulting in a topologically protected line of defects that can be easily moved with small forces (Fig. 6b). This feature enables the creation of a robust programmable memory~\cite{Corentin_Nature_2023}, but how to harness such topological frustration for on-demand shape changes with memory remains an intriguing open question.

\emph{Tessellations of unit cells based on anisotropic mechanism. ---} 
A global constraint can also be readily created by arranging tubes and filaments into circular plates. A notable example is that of anisotropic tubes arranged to form a circular plate~\cite{Baromorphs} (Fig. \ref{fig6}c). If the tubes deformed equally in all directions, the plate would remain flat. However, upon pressurization, the tubes elongate more in the circumferential direction than along the longitudinal one (see Equation (\ref{strain_tubes})). This anisotropic deformation cannot be accommodated by a flat plate, causing it to buckle out of plane to reach a 3D equilibrium shape that minimizes total elastic energy. For example, if the tubes are arranged radially, the plate expands more in the azimuthal direction than in the radial direction upon pressurization. This creates an excess angle in the plate, which destabilizes it into a shape with negative Gaussian curvature (Fig. \ref{fig6}c). Conversely, if the tubes are arranged in concentric circles, the plate expands more in the radial direction, triggering an instability that results in a conical shape (Fig. \ref{fig6}c).  Note that similar shape changes have also been observed in anisotropic plates comprising a dielectric elastomeric matrix   reinforced with stiff fibers \cite{Hajiesmaili_Lewis_Clarke_Sci_Adv_2022}. %\KB{here we need to reference papers. There are no references}%\KB{I would also add the examples of loop with rotating squares. Also the case of triangular lattice can go under loops. Do we know of another strategy does not involve loops?} %\cc{WHAT ABOUT THE SNAPPING DOMES OF DOUG HOLMES AND MATTEAO PEZZULA?}  

\begin{figure}[t!]
	\centering
	\includegraphics[width=0.98\textwidth]{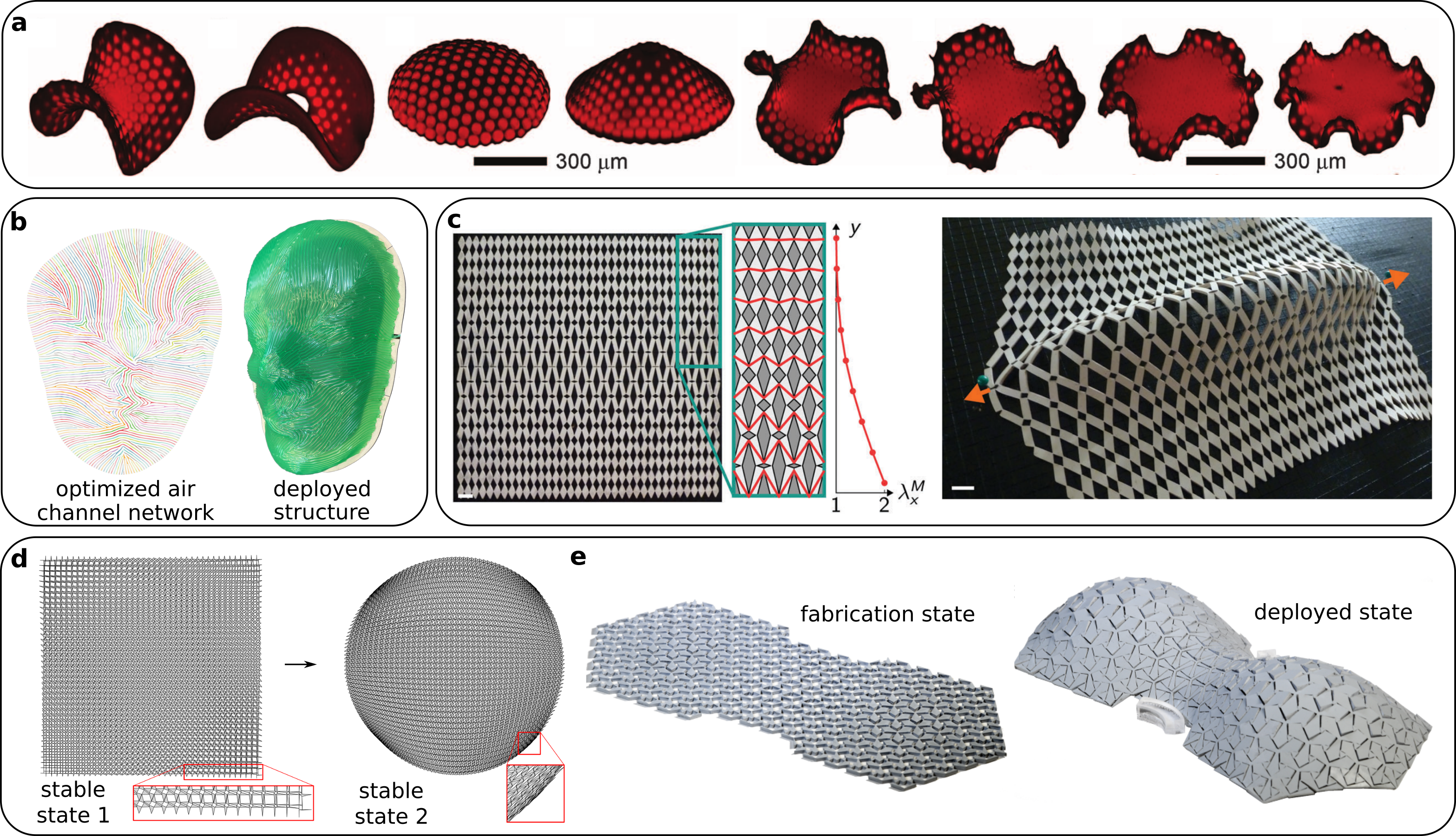}
	\caption{\textbf{Kinematically incompatible metamaterials made from non-periodic tessellations.}  {\bf a} Polymer gel plates patterned with varying swelling ratios buckle out of the plane in complex shapes that depend on the pattern~\cite{Kim_Hayward_Science_2012}. {\bf b} Inflatable plate made from non-periodic arragments of pneumatic tubes designed by computational optimization~\cite{Panetta_ACMTransGraphics2021}. The plate buckles out of the plane into the shape of the human face. {\bf c} Metamaterial composed of quads connected by their vertices and of spatially varying dimensions~\cite{Celli_2016}. The quads counter-rotate under stretch. The tiling expands to various degrees of strain across the metamaterial and hence leads to out-of-plane buckling under stretch. {\bf d-e} Metamaterials made from rotating unit deforming either {\bf d} in-plane~\cite{Peng_Plucinsky_Phys_Rev_Appl_2024} or {\bf e} out-of-plane~\cite{Tian_Chen_Mark_Pauly_ACM_2021} as a result of the kinematic incompatibility of their deformations. This kinematic incompatibility leads to an energy maximum the structure needs to overcome to deploy. As a result the deployed shapes are stable even when the loading is removed.}
	
	\label{fig7}
\end{figure}

\subsubsection{Non-periodic tesselations}

Periodic tessellations of kinematically incompatible unit cells   can only exhibit geometric frustration in the presence of loops. However, this constraint is unnecessary when using non-periodic tessellations, as frustration can also arise from mismatched strains across different regions of the tessellation. Early examples of this concept were demonstrated in polymer gel plates patterned with materials exhibiting varying swelling ratios~\cite{Kim_Hayward_Science_2012} (Fig.~\ref{fig7}a). Differential swelling creates in-plane strain variations across the plate, leading to kinematic incompatibilities that drive changes in Gaussian curvature and trigger out-of-plane buckling instabilities.

This approach has enabled complex shape changes in metamaterials made from both rotating and anisotropic unit cells. Metamaterials plates with spatially varying dimensions have been  designed  to induce strain differences across the sample that trigger  out-of-plane buckling instabilities and the formation of complex 3d morphologies ~\cite{Tian_Chen_Mark_Pauly_ACM_2021,Celli_2016}(Fig.~\ref{fig7}c). When the metamaterial is sufficiently thick, however, it deform in-plane. In this case, geometric frustration can lead to bistability, where both the initial and  deployed shape are stable~\cite{Choi_Dudte_Nat_mater_2019, Reza_Celli-arXiv_2024, Peng_Plucinsky_Phys_Rev_Appl_2024} (Fig.~\ref{fig7}d-e).  %Finally, out-of-plane deformation and bistability can be combined to create stable deployed shapes (Fig.~\ref{fig7}e). 

Similarly,  anisotropic unit cells with varying orientation and shape can be combined to form plates that experience spatially varying strains. Such strain variations in the plane   force changes in Gaussian curvature, causing the plate to buckle into a three-dimensional shape~\cite{Baromorphs, Griniasty_PRL2019, Griniasty_PRL2021, Panetta_ACMTransGraphics2021} (Fig.~\ref{fig7}b).

\section{Design Tools for achieving shape morphing.}

When designing shape-morphing metamaterials, the goal is often to create structures that can transform into specific, predefined target shapes. This design process can be framed as an inverse problem, where we aim to identify a metamaterial architecture capable of achieving the desired shape through mechanical deformation.

Robust and efficient algorithms such as topology optimization \cite{Sigmund_Maute_2013} have been developed to guide the design of structures with target responses in the linear regime. However, these methods are not directly applicable to the inverse design of shape-morphing mechanical metamaterials. This is due to the highly nonlinear nature of these systems, which often exhibit complex energy landscapes with multiple minima separated by large energy barriers, making them challenging to navigate. Consequently, solving the inverse problem for nonlinear metamaterials requires first selecting the design space (i.e. a set of unit cells and assembly rules). Once these are identified, computational tools can then be deployed to optimize the geometry of the unit cells, enabling the desired shape transformation.

\emph{Choosing the design space. ---}
The first step involves defining the design space and constraints. After selecting the type of unit cells and their tessellation pattern, a model must be developed to capture their behavior. There are two main approaches to modeling the response of shape-morphing metamaterials: kinematic constraints or finite energy deformations. In the kinematic approach, the elasticity of the metamaterial is disregarded, assuming it deforms according to an underlying rigid mechanism \cite{Choi_Dudte_Nat_mater_2019}. Metamaterials with a single underlying mechanism are ideal for robust shape-changing. Differently, mechanisms with multiple degrees of freedom allow for various target shapes, but are less robust and require more complex actuation \cite{Overvelde_deJong_Bertoldi_Nat_Commun_2016}.

However, it is important to note that fabricated metamaterials based on rotating units do not typically support idealized, mechanism-like behavior. Fabrication constraints, such as minimum hinge width, result in hinges with notable stiffness that dampens the desired mechanism-like response \cite{Meeussen_Bordiga_Bertoldi_Adv_Funct_Mater_2024}. In such cases, models relying solely on kinematics are insufficient, and fully elastic models must be employed, though these are generally more computationally demanding. To reduce computation costs, these models often simplify elastic problems by using plate~\cite{Panetta_ACMTransGraphics2021, Griniasty_PRL2021, Griniasty_PRL2019}, beam~\cite{Panetta_ACMTG_2019, Jennifer_Lewis_Boley_PNAS_2019}, or mass-spring~\cite{Yan_PNAS_2017, Rocks_PNAS_2017, Stern_PRX_2021} elements. Alternatively, detailed continuum models can be accelerated by surrogate models with AI~\cite{Bessa_AdvMat_2019,Alderete_NMI2022}, but the use of such data-driven models is not very well developed yet in the context of shape-morphing.

\emph{Computational design. ---} Once the design space has been selected, one can turn the design of a target shape-change into a computational optimization problem. The cost function is the distance between the actual and target shape change and the design parameters are the geometrical parameters of the metamaterial. If one is concerned with zero energy deformations, the optimization has then to be performed under constraints to ensure that the deformations remain at zero energy throughout the optimization~\cite{Dudte_Vouga_2016, Choi_Dudte_Nat_mater_2019, Kim_NatPhys2019,Choi_Dudte_Maha_2021_Phys_Rev_Res, Dudte_Choi_PNAS_2021, Dykstra_Coulais_arXiv_2023, Dudte_Maha_NCS_2023}. If one is concerned with finite energy deformations, one does not require such constraints, but the challenge is to have an efficient forward simulation to reduce computation time.

\section{Outlook}

In summary, this review has examined the different types of shape-morphing metamaterials developed to date and systematically identified the design strategies used. We conclude by highlighting key challenges for future research.  

\emph{Load-carrying capacity.} An obvious challenge is to create shape-morphing structures that can carry a load. Promising approaches are being introduced that use the concept of tensegrity via stiff fibers in tension~\cite{Siefert_Roman_Soft_Matter_2020}, granular contacts in compression~\cite{Fu_Jin_Adv_Funct_Mater_2019,Dreier2024beadedmetamaterials}, and the use of stiff materials that display plasticity~\cite{Liu_Nature2024}.
Yet systematic design remains hard. An active effort is currently taking place to meet this challenge notably by using multi-objective optimization methods~\cite{Panetta_ACMTransGraphics2021,Pauly_ACMTG2024} with exciting prospects for architecture.

\emph{Multiple shapes.} Most of the shape-morphing metamaterials either can morph into one shape only, or into multiple shapes but at the cost of complex actuation. %\cc{connect to multistability paragraph}
Despite early attempts using loading speed~\cite{Janbaz_Coulais_Nat_Commun_2024} or multistability~\cite{Anto_AdvFunctMat_2022}, creating multishape metamaterials with simple actuation remains a formidable challenge. The use of multistability discussed above throughout this paper is emerging as a particularly promising avenue to create metamaterials that can toggle between different stable shapes. As a result of this multistability, the state the metamaterial sits in depends on the loading history and in some cases many different states can be visited by suitably controlling the loading and unloading sequence. This approach is now routinely employed for mechanical computing~\cite{bense2021complex,yasuda2021mechanical,Kwakernaak_PRL2023,Liu_PNAS2024}, but remains largely open in the context of metamaterials for shape morphing~\cite{Anto_AdvFunctMat_2022}.
Another related challenge is how to perform the design of such multishape metamaterials. So far attempts to perform computational design of multishape metamaterials are able to create only relative simple shape-morphing that in addition are often muddied by spurious modes~\cite{Dieleman_van_Hecke_2019,Kim_NatPhys2019,Choi_Dudte_Maha_2021_Phys_Rev_Res,Dykstra_Coulais_arXiv_2023,vanmastrigt2024prospectingpluripotencymetamaterialdesign}.

\emph{Micron-scale fabrication.} An equally difficult challenge is to scale shape-morphing metamaterials down to the micron-scale. There the key challenge is the fabrication and the actuation. Although two-photon nanolithography can be used to create metamaterials with features as small as $150$nm, bulky instruments such as nanoindenters are required to make them shape-change. Another particularly promising approach is to nanofabricate responsive materials such as elastomers that can swell~\cite{Silverberg_Cohen_Nat_Mater_2015}, graphene bibeams~\cite{Miskin_PNAS2018,Miskin_Nature2020,Wang_Nature2022} and magnetic domains~\cite{Smart_Cohen_McEuen_Science_2024,Coulais_Groep_Science_2024}. Such metamaterials are always made by lithography or atomic layer depositions as two-dimensional origami or kirigami, thus the key challenge is to select the right combination of materials that offer compliance and responsiveness simultaneously and to design the folding out of the plane and the shape morphing once deployed. Those metamaterials offer compelling potential for applications in microrobotics in aqueous environments~\cite{Miskin_PNAS2018,Miskin_Nature2020,Wang_Nature2022} and for the control of light fields~\cite{Smart_Cohen_McEuen_Science_2024, Coulais_Groep_Science_2024}.

\emph{Robotics.} A fascinating area of application of shape-morphing metamaterials is that of robotics, for e.g. locomotion or gripping. Yet, for those robotic functionalities, not only does the metamaterial need to shape change, the shape-change is required to occur in a time-ordered fashion or a cycle, in order to do work on its environment. Although recent works have used solitons~\cite{Deng_Bertoldi_Sci_Adv_2020,During_soliton} or limit cycles~\cite{Brandenbourger2022limitcyclesturnactive} to achieve locomotion, how do design a cycle of shape changes rather than a unique shape change and how to control its dynamics remains a largely open problem.

\section*{Authors contributions}
C.C., K.B., and K.K.D. wrote the first draft of the article. K.K.D., C.C., and M.K. prepared the figures based on the concept proposed by K.B. and C.C. All authors reviewed and/or edited the article before submission and contributed to the discussion.

\section*{Acknowledgements}
K.K.D. acknowledges the support of the Polish National Science Centre (NCN) - project No. 2022/47/D/ST5/00280, and Polish Ministry of Science - program name: ‘Regional Excellence Initiative’, project No. RID/SP/0050/2024/1. C.C. acknowledges funding from the European Research Council under Grant Agreement No. 852587 and from the Netherlands Organisation for Scientific Research (NWO) under grant agreement VIDI 2131313.

\bibliographystyle{unsrt}
\bibliography{references2.bib} 

\end{document}